\documentclass[12pt,a4paper]{article}

\usepackage{graphicx}
\usepackage{amssymb,amsmath}
\usepackage[cp1251]{inputenc}  
\usepackage[T2A]{fontenc}
\usepackage[russian, english]{babel}
\usepackage[unicode]{hyperref}
\usepackage{indentfirst}

\title{On the concepts of vacuum and mass and the search for higgs}
\author{L.~B.~Okun\\
ITEP, Moscow, Russia}
\begin{document}
\maketitle

\bigskip
\begin{abstract}
Some recollections on the recent history of the concepts of vacuum and mass and the search for higgs.
According to the widely spread terminology the Higgs field permeates vacuum and serves as the origin of masses of all fundamental particles including the Higgs Boson --- the higgs.  
\end{abstract}

\subsubsection{Introduction}

This is a brief comment on the discovery of  ``a boson similar to Higgs Boson'', written on the request of Professor Phua and based  on personal recollections and on my recent book ``ABC of Physics: a very brief guide''. According to to the language of Feynman graphs accepted in this book the role of field is played by virtual particle.

















\subsubsection{Madison 1980 -- Bonn 1981}

I am starting my comment with two conferences about 30 years ago in Madison and Bonn.
In 1980 I was invited to give the summary talk at the 20th International Conference on High Energy Physics in Madison. I prepared the talk, but unfortunately was unable to participate in the conference as I was excluded from the final list of Soviet delegates. The extended text of the talk was published in May 1981 in the Uspekhi~\cite{ok1}.

In the section 7 ``Scalars'' of this article  I wrote: 
\begin{quote}
The experimental search for scalar particles is
a primary task... What seems very promising is the associated production of $Z$ and $H$ bosons: $e^+e^-\rightarrow Z^0H^0$. The search for this reaction is one of the most important tasks of the accelerator LEP.
\end{quote}

In August 1981 I gave a closing talk at the Lepton-Photon Symposium in Bonn. There I have chosen the search for higgs as the problem number one  of the high energy physics~\cite{ok2}.

\subsubsection{CERN 1981--1984}
At the end of 1981 I was elected a member of the CERN Science Policy Committee. In 1984 the decision was made to build in CERN the Large Hadron Collider\footnote{The term ``hadron'' was introduced in 1962 at the 11th International Conference on High Energy Physics~\cite{had}.}~\cite{lhc}.

\subsubsection{The books ``Leptons and Quarks'' 1981 and ``Particle Physics: The Quest for the Substance of Substance'' 1984}
 
The preparatory work on the talk at Madison was reflected in the Russian and English versions of the book ``Leptons and Quarks'' ~\cite{ok5, ok6},  where in particular basic formulas describing the production and decay of higgs were collected.

The essence of Madison and Bonn talks was semi-popularly presented in the Russian~\cite{ok7} and  English~\cite{ok8} versions of the book ``Particle Physics: The Quest for the Substance of Substance''. 

\subsubsection{The Higgs Hunter's Guide 1990}

An excellent review of predictions of various Higgs models was published in 1990~\cite{guide}. Among many hundreds of references
it contained the reference~\cite{ok5}, but not the references~\cite{ok2, ok8}. 

\subsubsection{Conference in Singapore  1990}

In the invited talk ``Beyond the Standard Model''~\cite{sin} at the 25th International Conference on High Energy Physics in Singapore I stressed the extension of the Higgs sector. 

\subsubsection{``The Relations of Particles'' 1991}

During the Singapore conference Professor K.K Phua invited me to publish with World Scientific a collection of 
my talks and review articles of the 1980's in a book ``The Relations of Particles''~\cite{rel}.
It begins with the 1981 talk in Bonn~\cite{ok2}. Then follow the opening talk ``Physics of High Energy --- 86'' at the 13th International Conference on High Energy Accelerators in Novosibirsk, the invited talk ``From pions to wions'' at the 20th International Cosmic Rays Conference in Moscow, 1987, the closing talk at the ``Neutrino --- '88'' in Boston, and the closing talk at the Symposium on the Fourth Family of Quarks and Leptons, Santa Monica, 1989. The book concludes with a  few articles on the concept of mass reprinted from ``Uspekhi'' and ``Physics Today'', 1989. 
   
\subsubsection{The Postcard from John Archibald Wheeler 1991 }

In 1979 on centennial birthday of Einstein his memorial sculpture was opened in Washington and a special postcard was issued. On one side of it is the sculpture of Einstein slightly covered by snow, on the other side --- the dedication words by Wheeler: 
\begin{quotation}
 The Albert Einstein Memorial

National Academy of Sciences, Washington, DC

``How can one most clearly say that science reaches beyond all national boundaries and belongs to all mankind?

                            <...>

Not by a pompous figure on a pedestal... (rather by) a figure around which young people can sit and think long, long thoughts; the figure who said, ``The ideals which have lighted my way, and time  after time have given me new courage  to face life cheerfully have been Kindness, Beauty and Truth'''' \emph{John Archibald Wheeler, Dedication of Einstein Monument, April 1979.}
\end{quotation}

On October 23, 1991 Wheeler has sent me this postcard with the following hand-written text:
\begin{quotation}
Dear Professor Okun --- Many thanks for your beautiful little World Scientific {\bf The Relations of Particles}. It makes the whole subject delightful - - and again upholds the correct idea of mass! Warm good wishes ---

John Wheeler

(cc. to W.Sci)
\end{quotation}

The two sides of the  Einstein postcard can be seen at~\cite{file94}. It would be interesting to look for the copy of this postcard in the archive  of the office of World Scientific.
 
\subsubsection{Another Postcard with Einstein Memorial}

Along with the postcard sent by Wheeler I have another postcard with the Albert Einstein memorial without snow~\cite{file96}. The Einstein equations can be clearly seen there, the last one is $E=mc^2$. This demonstrates the humorous side of the Wheeler's 1991 letter which in fact referred to the formula $E_0=mc^2$ in accord with the ``Dialog: Use and Abuse of the Concept of Mass'' of ref\cite{tw}.  
  
\subsubsection{From Galilei to Higgs 1991}

At the invitation of Professor Antonino Zichichi I gave the opening lecture ``The Problem of Mass: From Galilei to Higgs'' at the 1991 International School of Subnuclear Physics in Erice~\cite{gh}.  The school was dedicated to the famous equality of gravitational and inertial masses, traditionally ascribed to Galileo Galilei. I tried to explain that this equality is a myth, that there was no such concept as mass in the works of Galileo, and that there is only one concept of mass $E_0=mc^2$ now, four hundred years later.

\subsubsection{Electroweak Radiative Effects 1994}

I gave the closing lecture ``First Evidence for Electroweak Radiative Effects from  the Latest High Precision Data'' at the 1994 International School of Subnuclear Physics in Erice~\cite{ere}. The lecture was based on a series of articles published during the previous three years together with V.~A.~Novikov, M.~I.~Vysotsky, 
A.~N. Rozanov, V.~P.~Yurov, and N.~A.~Nekrasov on optimal parametrization of Electroweak Theory. 
   
\subsubsection{Dallas 1989 -- 1993}

In 1989 I was elected to the Scientific Policy Committee (SPC) of the Superconducting Supercollider Laboratory (SSC Lab) near Dallas. I was a member of SPC of SSC till the end of the SSC project in 1993. In 1993 together with two junior Russian physicists from CERN SPC I wrote a letter to President Clinton explaining to him why the construction of SSC should not be stopped.

\subsubsection{Marseille 1993}
 
At the International Europhysics Conference on High Energy Physics in Marseille I was invited to present the closing talk ``Outlook''~\cite{mars}. It ended with two sections: ``Quest for New Particles'' and ``Storm-Clouds''. I named higgs as the most necessary new particle,  and as the most serious storm-cloud --- the decision to stop the construction of the SSC, the flagship of HEP.

\subsubsection{Erice  1994}

In the invited talk ``Vacuum, Vacua: The Physics of Nothing'' at the International Conference on History of Original Ideas and Basic Discoveries in Particle Physics in Erice~\cite{harv} I briefly sketched the early history of several  ideas concerning possible properties of the vacuum. A revised version of this talk was presented seven years later
at Ann Arbor.

\subsubsection{Ann Arbor  2001}

An important role in the canonization of the Higgs boson was played by the conference ``2001: A Spacetime Odyssey'' held in Ann Arbor. The proceedings of this conference~\cite{odys} contain 17 invited talks.
At this conference Higgs gave the famous talk~\cite{higgs} ``My Life as a Boson: The Story of  `the Higgs' '' where he said:
\begin{quotation}
By 1976, when LEP was being planned, this had been introduced to experimentalists in ``a phenomenological profile of the Higgs boson'' by John Ellis, Mary K. Gaillard and Dmitri Nanopoulos. Apologizing for the vagueness of this profile, they concluded ``we do not wish to encourage big experimental searches for the Higgs boson, but we do feel that people performing experiments vulnerable to the Higgs boson should know how it may turn up''.      

Fifteen years later, with much of the Standard Model verified experimentally, John Gunion, Howard Haber, Gordon Kane, and Sally Dawson in ``The Higgs Hunter's Guide'' felt able to be more assertive: ``The success of the Standard Model has been astonishing. The central problem today in particle physics is to understand the Higgs sector.''

From 1989 onward measurements at LEP defined the parameters of the Standard Model with ever increasing precision.
\end{quotation}


At the conference in Ann Arbor I gave the talk ``Spacetime and vacuum as seen from Moscow''~\cite{ok9} with the following contents:  1.~Introduction 2.~Pomeranchuk on vacuum 3.~Landau on parity P and combined parity, CP 4.~Search and discovery of $K^0_L\rightarrow\pi^+\pi^-$ 5.~``Mirror world'' 6.~Zeldovich and cosmological term 7.~QCD vacuum condensates 8.~Sakharov and baryonic asymmetry of the universe, BAU 9.~Kirzhnits and phase transitions 10.~Vacuum domain walls  11.~Monopoles, strings, instantons, and sphalerons  12.~False vacuum  13.~Inflation 14.~Brane and Bulk.

\subsubsection{``Prestigious Discoveries at CERN'' 2003}

At the conference at CERN~\cite{cmr}, dedicated to 30th anniversary of the discovery of neutral currents and 20th anniversary of discovery of $W$ and $Z$ bosons I was invited to participate in the panel discussion on Future of Particle Physics. This conference was an important milestone on the road to the discovery of higgs.

\subsubsection{The article in AJP}

In 2008 I submitted a short note to American Journal of Physics in which I tried to summarize the situation with $E=mc^2$. The note was rejected several times  by the Editor, and was published only in 2009~\cite{ajp} under the title ``Mass versus relativistic and rest masses''. In one of the intermediate versions of this article I included a jocular short poem on $E=mc^2$ by Morris Bishop, which was suggested to me by Michael Gottlieb. It mocked:
\begin{verse}
But we know that energy equals mass \\
By the square of the speed of light. 
\end{verse}

\subsubsection{``The Energy and Mass in Relativity Theory'' 2009}

A book of 30 of my  published articles from 1968 to 2008 on various facets of the concept of mass  appeared in 2009~\cite{ok10}. The book begins with an article on the mass of the photon written jointly with I.~Yu.~Kobzarev, and ends with several articles on the concept of mass published in Uspekhi.

\subsubsection{Mumbai  2011} 

In 2011 in a concluding talk at the Lepton-Photon conference in Mumbai M. Peskin~\cite{pes} recalled the statement from my  1981 Madison talk that the discovery of the Higgs boson is the problem Number 1 of physics.

\subsubsection{``ABC of Physics''  2012}

Awaiting the results on higgs at LHC I published the Russian and English versions of my little book~\cite{ok11, ok12}. In it I tried to explain using the simplest high-school mathematics why it is so important to solve the problem of higgs and why it is necessary to explain to children the simple contemporary picture of the world. Only such educated children, when becoming grown-ups, will be able to solve the vital problems which face the mankind. 

\subsubsection{ ``ABC of Physics'' on Feynman diagrams}

A powerful instrument in physics is the method of Feynman graphs which can be applied at various scales. It is convenient to consider the free particles as real particles being on their mass shell. Particles off-mass shell are called virtual particles. The motion of real and virtual particles is depicted by lines, their interactions --- by vertices. The numerical coefficients of vertices are called coupling constants.

Note that our language is lagging behind our experience. Atoms (in greek ``not divisible'') are in fact divisible: they consist of electronic shells and nuclei, which consist of nucleons (protons and neutrons).  Not all elementary particles are elementary in the sense that they have no constituents: the nucleons as well as the other hadrons consist of more elementary (fundamental) particles --- quarks, bound by gluons. In the present language the quarks are as fundamental as the electrons and the other leptons. Though hadrons are not fundamental since they have constituents, they are usually called elementary as they are not divisible into quarks because of the property of confinement of gluons.  

\subsubsection{``ABC of Physics'' on gravity}

The Feynman graphs merge both relativity (with $c$) and quantum mechanics (with $\hbar$). In particular, the vertices
are determined by the spin of particles involved. In the Feynman graphs approach gravity caused by spin two gravitons is similar to all other interactions (for instance, electromagnetic) at low energies and momenta. It becomes singular only at Planck scale, where the consistent theory of gravity of 
pointlike particles is still missing.

The contemporary picture of the world is based on the concept of fundamental particles with integer (bosons) and half-integer (fermions) values of spin in units of $\hbar$. The interactions of particles are described by Feynman graphs. Thus according to ``ABC of Physics'' gravitons are included into the list of fundamental particles of the Standard Model.
In this respect ``ABC of Physics'' is radically different from the most of the talks and articles by other authors that commented on the discovery at LHC (see for instance ref \cite{wei}.




\subsubsection{``ABC of Physics'' and the term ``higgs''}

A remark on terminology is needed here. As seen from the book, I prefer a short term ``higgs'' with the lower case ``h'' to other names of this particle and especially to ``Higgs Boson''. This places the particle in the traditional line with other particles. 
    
\subsubsection{Epilogue and Prologue}

When this note was almost finished the discovery at LHC was reviewed  in Uspekhi~\cite{rub} and the first Russian translation of the 1981 Bonn talk was published ~\cite{ok13,ok14}.  
Conference ``Higgs Hunting 2012'' took place in France, it was summarized by M. E. Peskin~\cite{pes1}. 
A talk on LEP-3 was given at CERN by Patric Janot. The file of this talk \cite{lep3} was kindly sent to me by Ilya Ilyich Zuckerman.                                                         

\end{document}